\documentclass[12pt]{article}

\usepackage{times}
\usepackage{geometry}
\usepackage[utf8]{inputenc}
\usepackage{enumitem,amssymb}
\usepackage{ragged2e}
\newlist{thematic}{itemize}{8}
\setlist[thematic]{label=$\square$}
\usepackage{pifont}

\usepackage{natbib}
\usepackage{setspace,wrapfig}
\usepackage{mdframed}
\usepackage{mathptmx} 
\usepackage{microtype}
\usepackage{graphicx}
\usepackage[T1]{fontenc}
\usepackage[svgnames]{xcolor}
\usepackage{mathtools}
\usepackage{bbding}

\usepackage[pages=some]{background}
\backgroundsetup{contents=\includegraphics[]{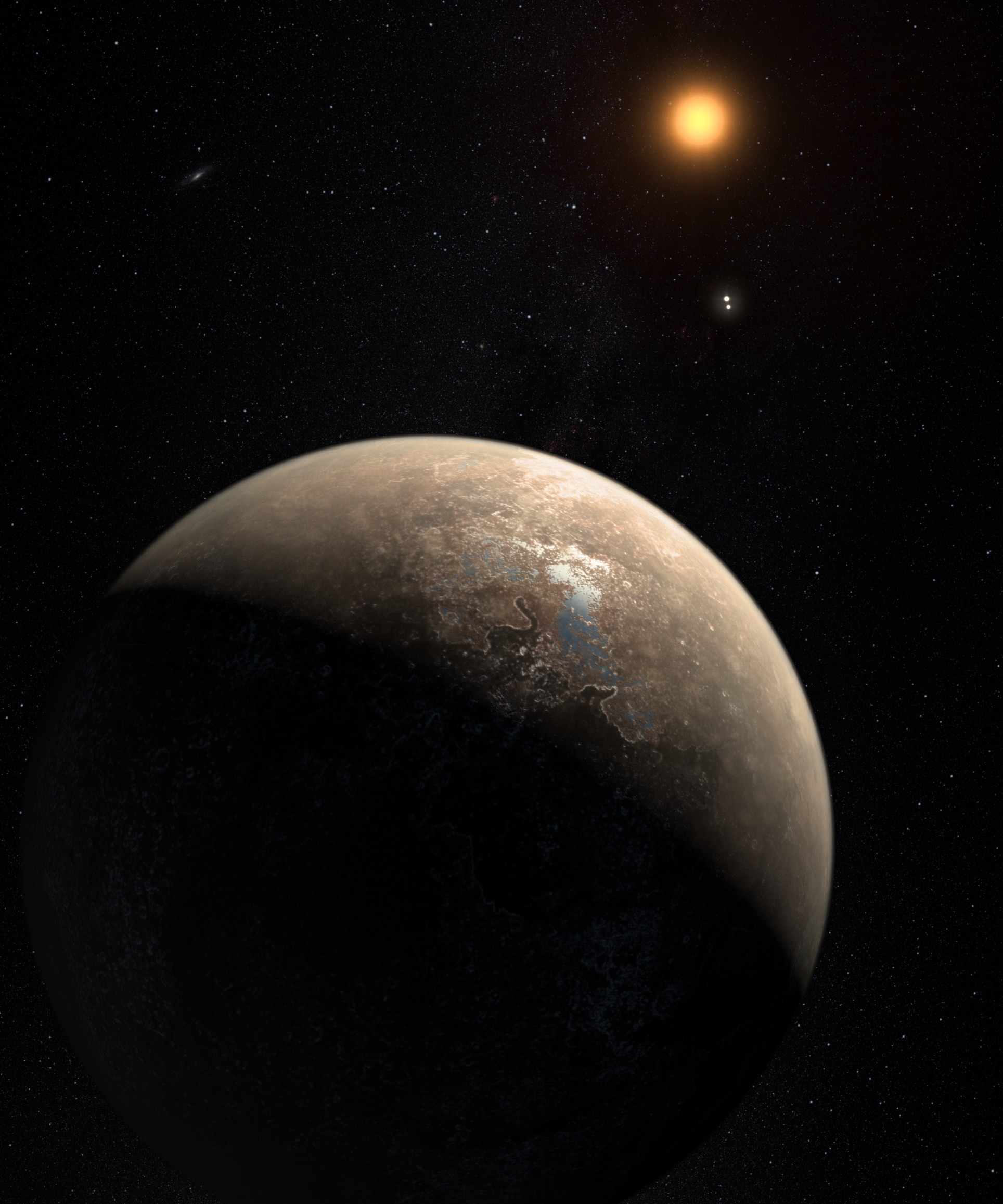},scale=0.6,opacity=0.3,vshift=-300pt}

\begin{document}
\raggedright

\BgThispage
\huge 
Astro2020 Science White Paper \linebreak

\vspace{-0.2in}
High-Energy Photon and Particle Effects on Exoplanet Atmospheres and Habitability \linebreak
\normalsize

\vspace{-0.05in}

\noindent \textbf{Thematic Areas:} \hspace*{60pt} \rlap{$\square$}{\CheckmarkBold} Planetary Systems \hspace*{10pt} \rlap{$\square$}{\CheckmarkBold} Star and Planet Formation \hspace*{20pt}\linebreak
$\square$ Formation and Evolution of Compact Objects \hspace*{31pt} $\square$ Cosmology and Fundamental Physics \linebreak
  \rlap{$\square$}{\CheckmarkBold} Stars and Stellar Evolution \hspace*{1pt} $\square$ Resolved Stellar Populations and their Environments \hspace*{40pt} \linebreak
  $\square$    Galaxy Evolution   \hspace*{45pt} $\square$             Multi-Messenger Astronomy and Astrophysics \hspace*{65pt} \linebreak

\vspace{-0.06in}
  
\textbf{Principal Author:}

Name:	Jeremy J. Drake
 \linebreak						
Institution:  Center for Astrophysics \vrule ~Harvard \& Smithsonian
 \linebreak
Email: jdrake@cfa.harvard.edu
 \linebreak
Phone:  617 496-7850
 \linebreak

\vspace{-0.1in}

\textbf{Co-authors:} 
Juli\'an D. Alvarado-G\'omez (CfA),
Vladimir Airapetian (NASA/GSFC),
Costanza Argiroffi (Dip. di Fisica e Chimica\ Univ.\  Palermo),
Matthew K. Browning (Univ.\ Exeter),
Damian J. Christian (Cal.\ State Northridge),
Ofer Cohen (UMass Lowell),
Lia Corrales (Wisconsin, Madison),
William Danchi (NASA/GSFC),
Miguel de Val-Borro (NASA/GSFC),
Chuanfei Dong (Princeton), William Forman (CfA), 
Kevin France (LASP, Univ.\ Colorado),
Elena Gallo (Univ.\ Michigan), 
Katherine Garcia-Sage (NASA/GSFC),
Cecilia Garraffo (Harvard), 
Dawn M. Gelino (NASA Exoplanet Science Institute),
Guillaume Gronoff (NASA/LaRC),
H. Moritz G\"{u}nther (MIT),
Graham M. Harper (CASA, Univ.\ Colorado),
Rapha\"{e}lle D. Haywood (CfA),
Margarita Karovska (CfA), 
Vinay Kashyap (CfA), 
Joel Kastner (RIT), Jinyoung Serena Kim (University of Arizona), 
Maurice A. Leutenegger (NASA/GSFC \& CRESST/UMBC),
Jeffrey Linsky (JILA, Univ.\ Colorado), Mercedes L\'opez-Morales (CfA),
Giusi Micela (INAF-Oss.\ Astron.\ Palermo),
Sofia-Paraskevi Moschou (CfA),
Lidia Oskinova (Univ.\ Potsdam),
Rachel A. Osten (STScI, JHU),
James E. Owen (Imperial), 
Katja Poppenhaeger (Univ.\ Potsdam),
David A. Principe (MIT),
John P. Pye (Univ.\ Leicester),
Salvatore Sciortino (INAF-Oss.\ Astron.\ Palermo),
Panayiotis Tzanavaris (NASA/GSFC \& CRESST/UMBC), 
Brad Wargelin (CfA),
Peter J. Wheatley (Univ.\ Warwick),
Peter K. G. Williams (CfA), 
Elaine Winston (CfA), 
Scott J. Wolk (CfA)
P. Wilson Cauley (LASP, Univ.\ Colorado)
  \linebreak

\justifying
\vspace{-0.05in}

\noindent \textbf{Abstract:}
It is now recognized that energetic stellar photon and particle radiation evaporates and erodes planetary atmospheres and controls upper atmospheric chemistry. Key exoplanet host stars will be too faint at X-ray wavelengths for accurate characterization using existing generation and future slated X-ray telescopes.  Observation of stellar coronal mass ejections and winds are also beyond current instrumentation.  In line with the {\it Committee on an Exoplanet Science Strategy} recognition that holistic observational approaches are needed, we point out here that 
a full understanding of exoplanet atmospheres, their evolution and determination of habitability requires a powerful
high-resolution X-ray imaging and spectroscopic observatory.  This is the only capability that can: (1) characterize by proxy the crucial, difficult to observe, EUV stellar flux, its history and its variations for planet hosting stars; (2) observe the stellar wind; (3) detect the subtle Doppler signatures of coronal mass ejections.  

\pagebreak
\section{What Conditions Control Exoplanet
Habitability?} 


The rate at which gas is lost from an exoplanet's atmosphere is
critical for the survivability of surface water. Atmospheric mass loss can
be driven by both thermal and non-thermal processes, which
depend upon the radiation and winds of their host stars. The dominant
thermal process is hydrodynamical outflow energized by extreme ultraviolet
(EUV; 100--912~\AA) and X-radiation (0.1--100~\AA) that heat the exoplanet's
thermosphere and levitate gas against the exoplanet's gravitational
potential \citep[e.g.][]{Owen.Jackson:12}.
Photodissociation and ionization of molecules, including
water and CO$_2$, by the
stellar UV and EUV radiation increases the mass-loss rate by
producing lighter atoms (e.g., H) that are more easily lost to space. Most of
the thermospheric heating is by EUV photons but this radiation 
cannot be observed directly because of interstellar H absorption.  The chromospheric UV and FUV are inadequate EUV proxies.  The strength and
spectral energy distribution of a star's EUV emission instead arises from the transition region and corona. The 30--60~\AA\ range contains many of the same ionization stages that are important in the EUV range.  Observing these enables prediction of the EUV spectrum. 
Detecting the relevant lines in exoplanet hosts requires 
a high-resolution ($R\geq 5,000$) spectrum that is not feasible with {\em any existing or slated future missions}, including {\it Chandra, XMM-Newton} or {\it ATHENA}.  

\begin{wrapfigure}{R}{10cm}
\vspace{-0.8cm}
\begin{mdframed}[backgroundcolor=MidnightBlue!20] 
\vspace{-0.22in}
\section*{What type of mission is needed?} 
\vspace{-0.13in}
 \sf  A flagship X-ray space telescope with the following:\\
$\bullet$~~$\times$50 more effective area than {\it Chandra}
\\
$\bullet$~~Grating spectrometer with resolving power $R\geq 5000$\\
$\bullet$~~A microcalorimeter with 3~eV or better resolution
\end{mdframed}
\vspace{-0.7cm}
\end{wrapfigure}

The irradiation history of a planet also depends on the host star's rotation rate, faster rotators producing larger radiation doses over time by an order of magnitude or more than slower rotators \citep{Johnstone.etal:15}.  To understand the range and likely radiation doses, it is essential to map out the EUV radiation through time for stars of similar ages but different rotation rates.  This requires observations of open clusters with known ages at high spectral resolution in the soft X-ray range (30--100~\AA) and a facility with with effective area of about  50$\times$ that of {\it Chandra}.

\begin{figure}[ht]
\includegraphics[width=0.51\textwidth]{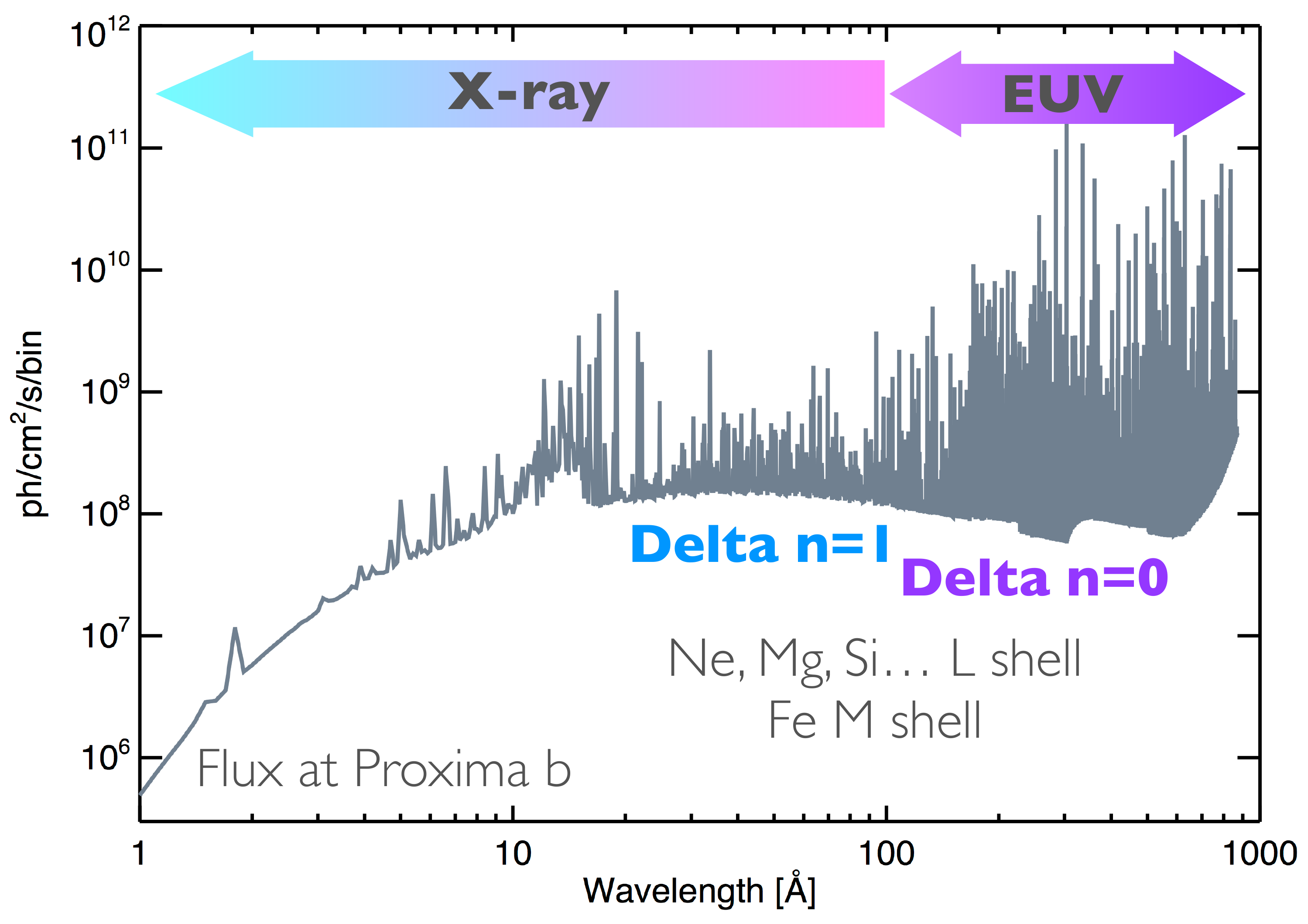} 
\includegraphics[width=0.49\textwidth,angle=0]{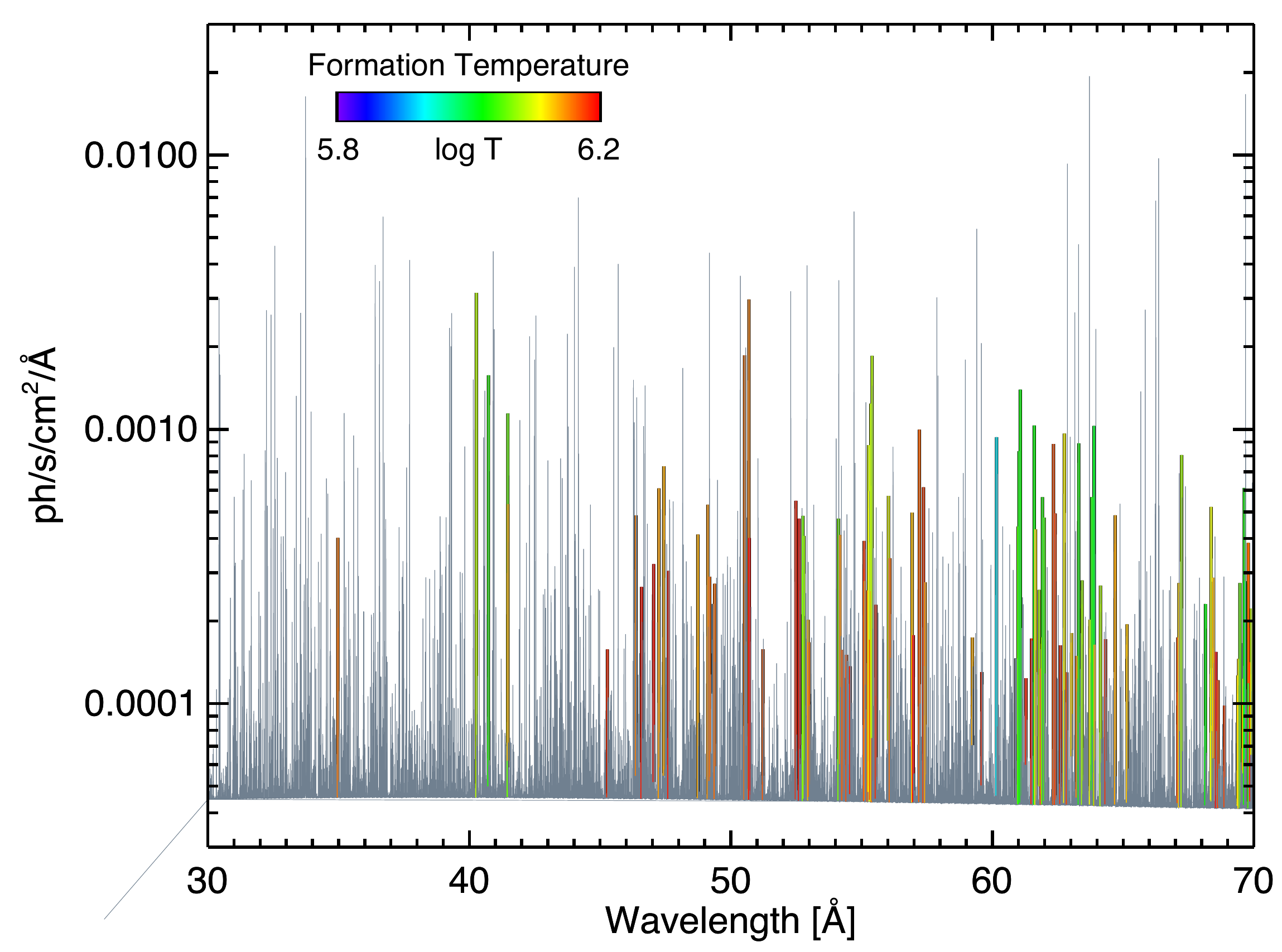}
\caption{\sl Left: The key X-ray to EUV spectral region computed for Proxima 
Centauri b
and responsible for upper planetary atmospheric ionization, heating and loss.  Coverage at high spectral resolution in soft X-rays is essential for understanding the EUV emission: The 30-60~\AA\ range exhibits 
transitions of the same ions that dominate the shorter EUV wavelengths. 
Right: The soft X-ray range, highlighting in colour lines formed at temperatures below $\log T=6.2$ that could be observed by a sensitive soft X-ray grating spectrometer and used to measure by proxy the EUV flux. }
\label{f:x-ray-euv}
\end{figure}

The X-ray emission of stars is variable on many time scales
especially for M dwarfs, which many astronomers think are the best host star candidates for locating nearby habitable exoplanets.  
Young rapidly-rotating stars
have high X-ray and EUV emission and emit energetic flares. Long-duration
monitoring of the optical radiation of G-type stars by {\it Kepler} shows
that high-energy superflares (total energy $E>10^{32}$ ergs) are
likely on a time scale of $\sim 500$ days for slowly rotating 
solar-like stars but are far more common on young G-type stars, and 
occur as often as 1 per 10 days \citep{Shibayama.etal:13}. 
Superflares have been observed with energies as large as
$E=10^{35}$ ergs. {\it Chandra} has observed superflares on M dwarf and young
stars, but the high-resolution spectra of superflares and also of more modest
flares needed to infer their EUV emission require a high resolution large effective area soft X-ray spectrometer 
(see Figure~\ref{f:x-ray-euv}).

\section{Stellar Winds and Exoplanet Atmospheric Loss}

The flow of ionized stellar wind electrons and protons erode an
exoplanet's atmosphere. Ions produced by
photoionization or charge-exchange reactions in the outer
atmospheres of exoplanets can be picked up by the magnetic field in
the stellar wind and expelled, can be lost through a ``polar wind''.
Simulations show that
such wind- and photoionization-driven processes can be a very important mass-loss agent for Earth-like planets around M stars
\citep{Garraffo.etal:16,Dong.etal:17,Garcia-Sage.etal:17,Airapetian.etal:17}. 
Recent measurements by the MAVEN satellite \citep{Brain.etal:16}
confirm
previous estimates that the primary mass-loss mechanism for water on
Mars is erosion by the solar wind.

The mass loss rates for late-type dwarfs are extremely difficult to measure as the solar mass-loss rate is only about
$1.5\times 10^{-14} M_{\odot}$~yr$^{-1}$. Radio observations yield only upper limits. There are indirect estimates of
mass-loss rates up to 100 times larger for four G and K stars with stronger
magnetic fluxes than the Sun based on Ly$\alpha$ absorption in the ``wall" of hydrogen at the stellar analogy of the heliopause \citep{Wood.etal:14}.
There are only two estimates using this technique of mass-loss rates for M stars---$8 \dot{M}_\odot$ for the active M3.5 dwarf EV~Lac and an upper limit of $< 10 \dot{M}_\odot$ for Proxima.

There is a clear need for new techniques for measuring the winds of a much larger sample of exoplanet host stars.  Such a technique is enabled by sensitive, high spatial resolution X-ray imaging.

The ionized stellar wind interacts with neutral atoms in the ISM and the astrosphere through radiationless collisional transfer of one or sometimes multiple electrons from a neutral ISM atom or molecule to a wind ion. Electrons captured into the upper levels of highly ionized metals cascade to lower levels, emitting X-rays.  
The resulting X-ray spectrum is dominated by K-shell emission from H-like and He-like ions of C, O, N, and Ne. The conversion to wind mass loss rate is direct.   
An attempt by \cite{Wargelin.Drake:02} to detect the charge exchange wind signature of Proxima using {\it Chandra} observations yielded only an upper limit of $3\times 10^{-13}M_\odot$~yr$^{-1}$.  Sub-arcsecond spatial resolution, high sensitivity and low background are required to make detections.
With new detector technology such as the X-ray  microcalorimeter, a sensitive next generation X-ray mission with arcsecond or better imaging will be able to observe the charge exchange signatures of stars out to at least 10pc for solar-like mass loss rates, and to larger distance for higher rates, enabling winds to be mapped out with stellar activity level and spectral type and generally applied to exoplanet systems.

Coronal plasma that is not confined by strong
magnetic fields must participate in the stellar wind expansion. 
A mission with high soft X-ray resolution
reaching $\lambda/\Delta \lambda =5000$, corresponding to 60 km~s$^{-1}$, and the possibility of measuring
flow velocities three times smaller for bright emission lines, 
will also have the capability to measure stellar winds directly.  This would be
totally new science that only a large area, high resolution X-ray mission could accomplish.


\section{Coronal Mass Ejections}

Strong X-ray flares on the Sun are usually accompanied by the ejection
of cooler material (roughly 10,000~K) that had previously been
confined by magnetic fields that became disrupted during the
flare. The ejected material, generally called coronal mass ejections
(CMEs), may also contain high energy protons accelerated in the
flare and CME shock front. CMEs differ from the quasi-steady solar wind in two respects:
they are orders of magnitude denser, and are spatially confined.

\citet{Segura.etal:10} modeled the effect of
a superflare ($E\approx 10^{34}$ erg) and CME impact on a hypothetical 
Earth-like exoplanet located in the habitable zone (0.16 AU) of the 
flare star AD Leo (dM3e). High energy protons with energies greater 
than 10 MeV severely depleted nitrogen oxides, and subsequently ozone, 
in the atmosphere for 2 years.  \citet{Airapetian.etal:16} found CME energetic particles can create important prebiotic molecules and alter atmospheric greenhouse gases potentially important for the Faint Young Sun paradox.

\begin{figure}[ht]
\includegraphics[width=0.52\textwidth]{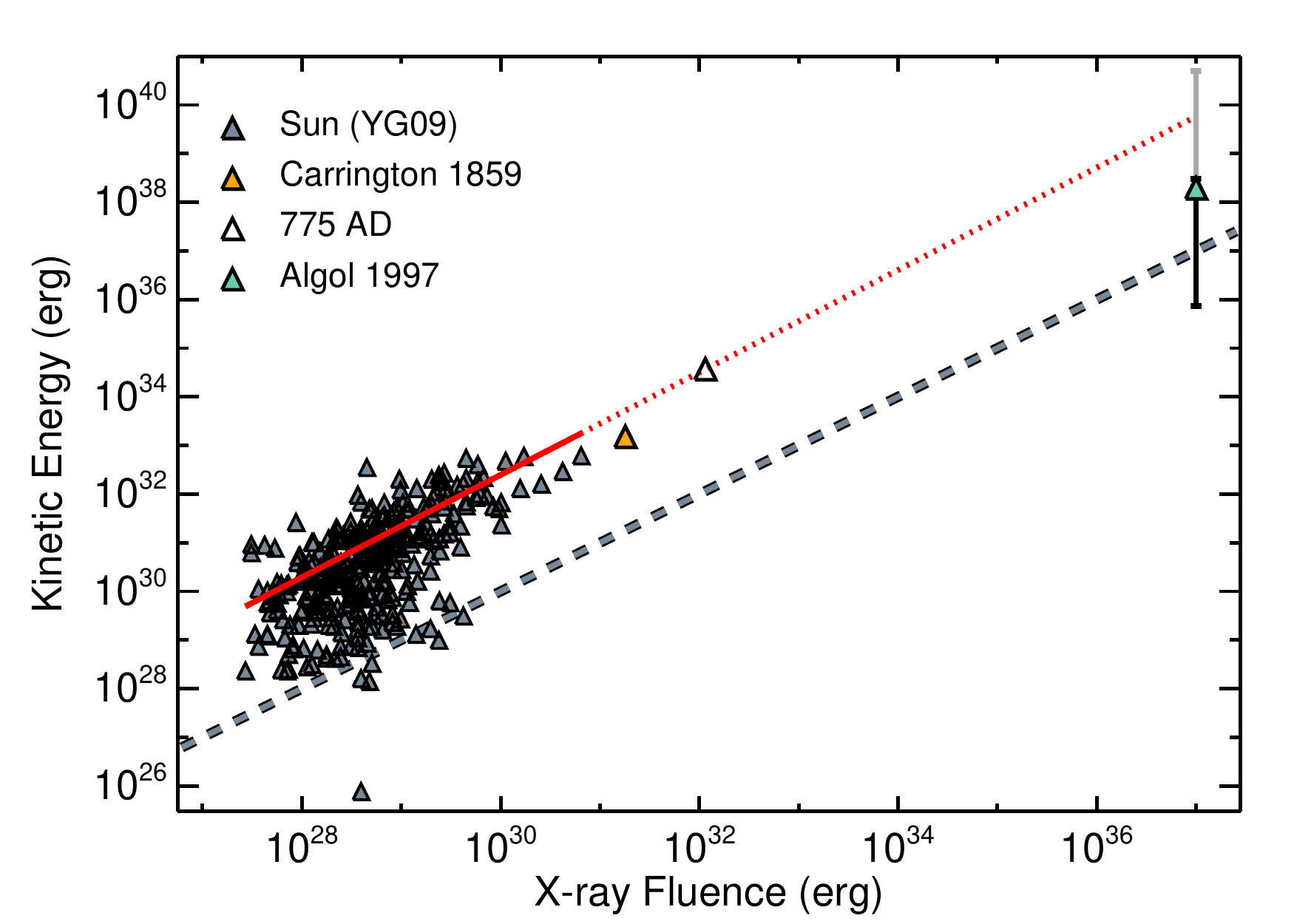}
\includegraphics[width=0.48\textwidth]{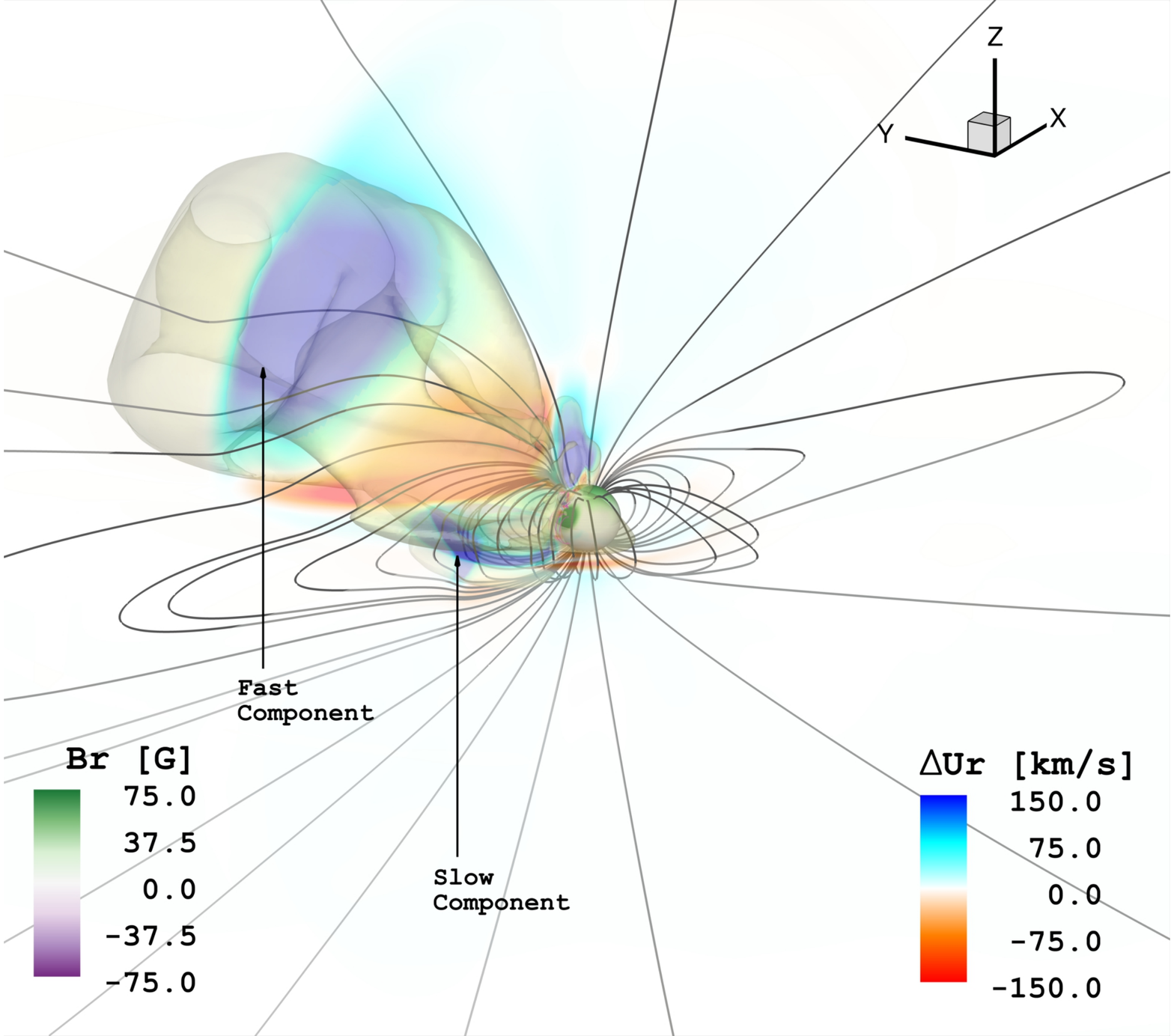} 
\caption{\sl Left: The kinetic energy vs.\ associated flare X-ray fluence for solar CMEs and a large CME and flare event on Algol \citep{Moschou.etal:17}. 
Extrapolating the relation to large events on more active stars is extremely uncertain, requiring definitive CME detections and measurements for characterization.
Right: An MHD CME simulation for a moderately active solar-like star \citep{Alvarado-Gomez.etal:18}. Plasma is compressed and accelerated outward by the CME front, yielding observable Doppler shifts, $\Delta U_r$,  of up to 100~km~s$^{-1}$ or so.  Such shifts would be detectable with a sensitive large area grating spectrometer. For effective areas $50\times$ that of {\it Chandra}, CMEs on active stars out to 200~pc and inactive stars to 20~pc could be detected.
}
\label{f:cmes}
\end{figure}

These studies demonstrate the acute need for observations of stellar CME events.  No such events have been definitively detected,
although there are searches underway at low frequency radio wavelengths. Extrapolations of solar CME-flare relationships (Figure~\ref{f:cmes}) are uncertain by orders of magnitude but are sorely needed to understand what CME activity exoplanets are experiencing.
High-energy protons are very difficult to observe, but the cooler material
in stellar CMEs, or the associated compression wave in the corona, should be observable by a sensitive high resolution X-ray spectrometer.  
There are two X-ray detections of probable CMEs where the cool, dense material is seen in absorption as it passes in front of the flaring corona: The 20 August 1980 flare on Proxima Cen observed by
{\em Einstein} \citet{Haisch.etal:83}; and the 30 August 1997 superflare on Algol 
observed by BeppoSAX \citep{Moschou.etal:17}. 

High-resolution spectroscopy at X-ray wavelengths
could routinely and definitively observe the tell-tale Doppler shifts of CMEs or their coronal compression waves (Figure~\ref{f:cmes}) and identify their physical properties, including their thermal structure, masses and energies. 
A combination of
high throughput and high spectral resolution will be critical, mapping out CME frequency and energy vs optical and X-ray flare diagnostics for exoplanet hosts directly, and generally as a function of spectral type and activity level.


    
\section{Transmission spectroscopy of exoplanet atmospheres}

X-rays are powerful diagnostics of planetary upper atmospheric gas density structure and chemical composition. 
The transit of the hot Jupiter HD189733b was detected through X-ray absorption by oxygen in {\it Chandra} observations by \citet{Poppenhaeger.etal:13}, who found the scale height of X-ray absorbing gas was higher than suggested by optical and UV transits.  
Hot Jupiters and similar giant close-in planets are important for improving theory and models describing atmospheric loss.   


\begin{figure}[ht]
\includegraphics[width=0.52\textwidth]{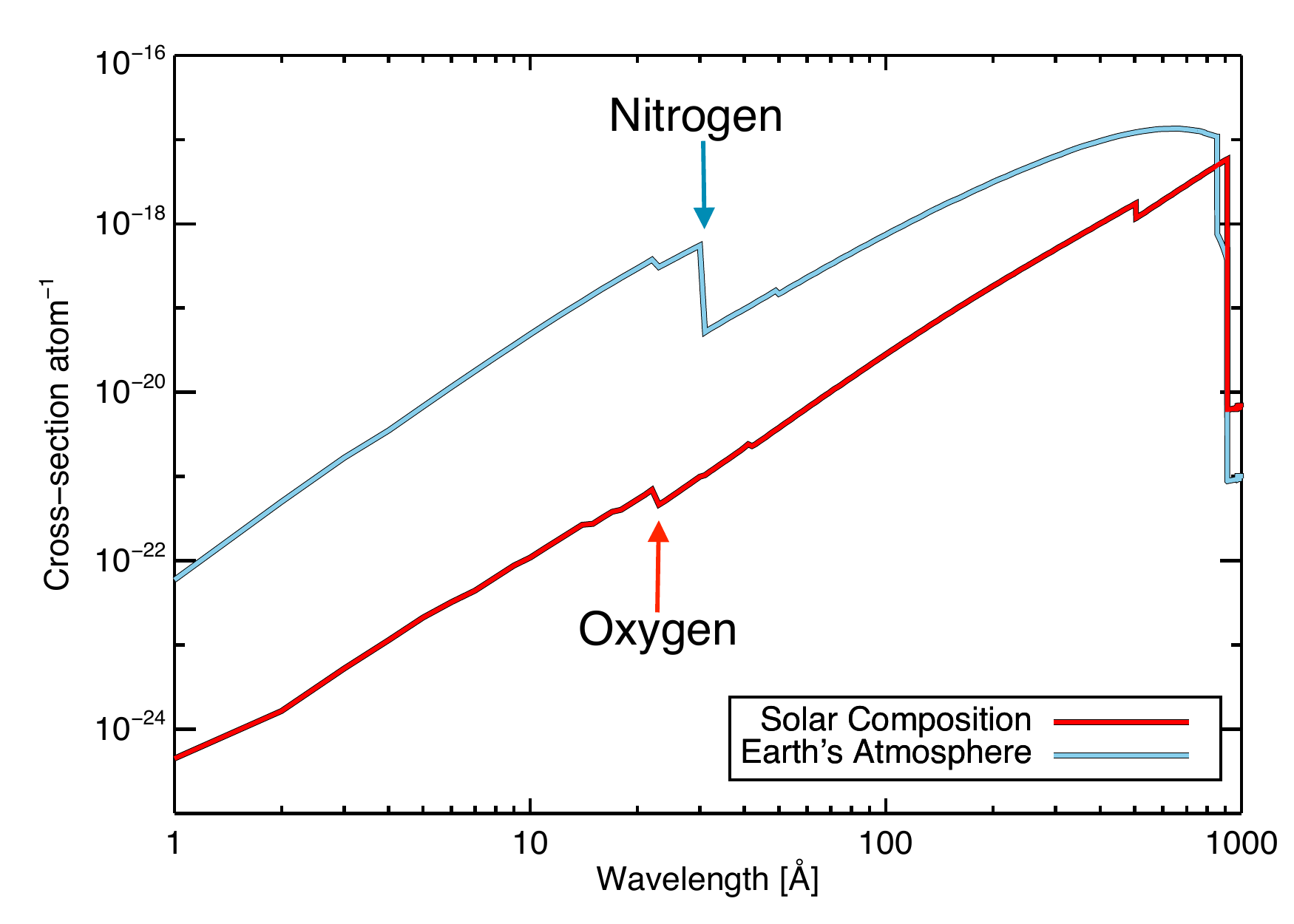} 
\includegraphics[width=0.48\textwidth]{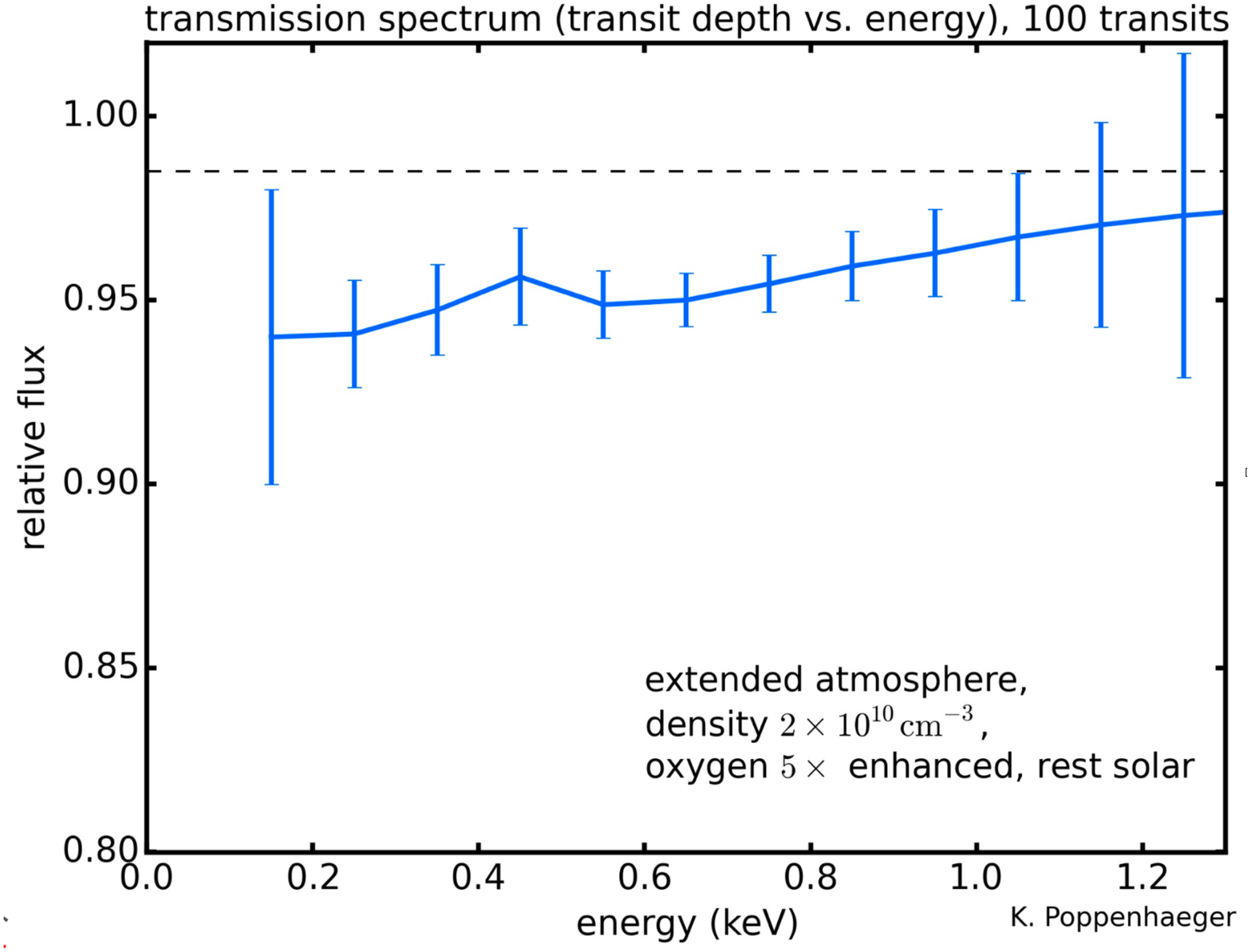}
\caption{\sl A large area X-ray observatory will be able to measure gas bulk composition from transmission spectroscopy. Left: Illustration of the enormous difference in X-ray transmittance of gas with solar and Earth's atmosphere compositions.  Right: Simulation of detection of the 0.5~keV oxygen absorption edge betraying enhanced O abundance for 100 transits of a superearth planet around an M dwarf for a telescope with $50\times$ the area of {\it Chandra} (by K. Poppenhaeger).}
\label{f:absorb}
\end{figure}

X-ray absorption measures gas {\em bulk chemical composition} (Figure~\ref{f:absorb}) along the line-of-sight---in this case in the transiting exoplanet atmosphere backlit by the host star's corona. 
Such measurements are unique to the X-ray range, but only the very closest hot Jupiters are accessible with {\it Chandra} and {\it XMM-Newton}, and then only at low signal-to-noise ratio.  
An observatory with an effective area $50\times$ that of {\it Chandra} will be able to observe HD 189733b-like transits out to 140 pc, a factor of more than 300 improvement in survey volume over current missions.  Combination with optical/IR data will provide a powerful probe for clouds and hazes that can confuse IR spectroscopic analyses \citep{Sing.etal:16}.
By coadding observations of many transits, such a mission could also open studies to larger habitable planets, such as super Earths around nearby M dwarfs (Figure~\ref{f:absorb}).

\section{Summary}

Exoplanet atmospheric loss and evolution cannot be properly understood without a powerful X-ray observatory capable of high spectral resolution of $R \geq 5,000$ at soft X-ray wavelengths, a microcalorimeter for higher energy high resolution imaging  spectroscopy, a large effective area at least several decades greater than that of {\it Chandra}, and with spatial resolution better than 1~arcsecond.

\pagebreak

\small
\begin{spacing}{0.3}
\bibliographystyle{aasjournal}

\bibliography{nas_refs}
\end{spacing}

\end{document}